\documentclass[%
 reprint,
 superscriptaddress,
 amsmath,amssymb,
 aps,
 floatfix,
]{revtex4-2}

\usepackage{eqnarray}
\usepackage{amsthm}
\usepackage{fancyhdr}
\usepackage{subfigure}
\usepackage{epsfig}
\usepackage{bbm}
\usepackage{subfigure}
\usepackage{bm}
\usepackage{float}
\usepackage{xcolor}

\usepackage{graphicx}
\usepackage{dcolumn}
\usepackage{bm,braket}

\usepackage{soul}

\begin{document}

\preprint{}

\title{Migration and separation of polymers in non-uniform active baths}

\author{Pietro Luigi Muzzeddu}
\affiliation{SISSA - International School for Advanced Studies, via Bonomea 265, 34136 Trieste, Italy}%
\author{Andrea Gambassi}
\affiliation{SISSA - International School for Advanced Studies, via Bonomea 265, 34136 Trieste, Italy}
\affiliation{INFN, Sezione di Trieste, Trieste, Italy}
\author{Jens-Uwe Sommer}
\affiliation{Leibniz-Institut f\"ur Polymerforschung Dresden, Institut Theory der Polymere, 01069 Dresden, Germany}
\affiliation{Technische Universit\"at Dresden, Institut f\"ur Theoretische Physik, 01069 Dresden, Germany}
\author{Abhinav Sharma}
\affiliation{Faculty of Mathematics, Natural Sciences, and Materials Engineering: Institute of Physics, University of Augsburg, Universit\"atsstra{\ss}e 1, 86159 Augsburg, Germany}
\affiliation{Leibniz-Institut f\"ur Polymerforschung Dresden, Institut Theory der Polymere, 01069 Dresden, Germany}

\date{\today}

\begin{abstract}
Polymer-like structures are ubiquitous in nature and synthetic materials. Their configurational and migration properties are often affected by crowded environments leading to non-thermal fluctuations. 
Here, we study an ideal Rouse chain in contact with a non-homogeneous active bath, characterized by the presence of active self-propelled agents which exert time-correlated forces on the chain. By means of a coarse-graining procedure, we derive an effective evolution for the center of mass of the chain and show its tendency to migrate towards and preferentially localize in regions of high/low bath activity depending on the model parameters. In particular, we demonstrate that an active bath with non-uniform activity can be used to separate efficiently polymeric species with different lengths and/or connectivity.
\end{abstract}

\maketitle

Living systems continuously exchange information and energy with the surrounding environment and their biological function unavoidably relies on mechanisms that are only allowed out of equilibrium~\cite{Fang2019nonequilibrium}. This inherent nonequilibrium state, exemplified by the hallmark feature of self-propulsion, gives rise to a diversity of collective behaviors shared by biological systems across various scales, ranging from molecular motor assemblies~\cite{Guerin2010coordination, Holzbaur20104coordination, Julicher1995cooperative} to swarming bacteria~\cite{kearns2010field} and flocking birds~\cite{toner1995long, toner1998flocks, toner2005hydrodynamics}. While a comprehensive theory that encompasses the diverse properties of living matter is still elusive due to the astonishing complexity of the biological world, significant efforts have been directed towards constructing a theoretical framework for \emph{active matter}~\cite{marchetti2013hydrodynamics, julicher2018hydrodynamic, gompper2020motile, fodor2016far, demagistris2015introduction}. Prominent examples from biology include flagellated bacteria~\cite{berg2004coli}, algae~\cite{polin2009chlamydomonas, geyer2013cell} and other motile microorganisms~\cite{elgeti2015physics}, molecular motors on cytoskeletal filaments~\cite{howard1996movement}, active worms~\cite{deblais2020phase, deblais2020rheology} and many others. 
Active colloidal molecules are also experimentally synthesized in the lab~\cite{elgeti2015physics, lowen2018active, bechinger2016active} using techniques such as self-diffusiophoresis via catalytic reactions~\cite{howse2007self, valadares2010catalytic, theurkauff2012dynamic}, light-induced self-thermophoresis~\cite{jiang2010active}, nonreciprocal deformation cycles~\cite{dreyfus2005microscopic, najafi2004simple}, and the integration of biological components with synthetic structures in biohybrid systems~\cite{williams2014self}.\\
\begin{figure}
    \centering
    \includegraphics[width=0.45\textwidth]{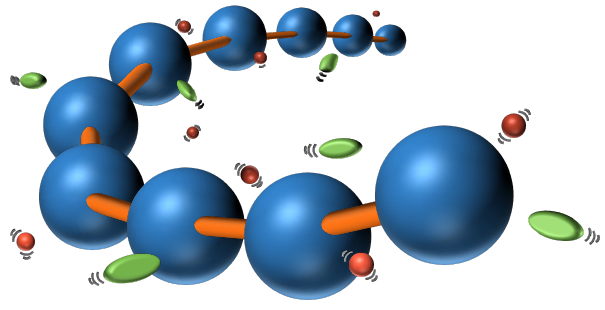}
    \caption{%Schematic bead-and-stick representation
    Sketch of a linear %polymeric molecule 
    polymer immersed in an active bath. The blue monomers, connected by orange %internal 
    bonds, interact with the red passive molecules (thermal bath) and are subjected to additional time-correlated forces due to the collision with the green active agents suspended in the surrounding fluid. These nonequilibrium interactions %bring in nonequilibrium fluctuations which 
 affect both the conformational and migration %properties
 statistics of the polymer.}%tracer chain. }
    \label{fig:sketch}
\end{figure}
Numerous active biological systems, including some of the examples mentioned above, appear as filamentous or polymer-like structures. It is well-established that several polymeric molecules in the interior of a cell rely on a variety of active reactions to regulate their biological functions. For example, DNA is continuously processed by enzymes such as DNA-polymerase and helicase to ensure its successful replication~\cite{alberts2017molecular}, ribosomes slide along RNA strands to synthesize proteins~\cite{alberts2017molecular} and the chromosomal loci dynamics is strongly affected by ATP-dependent non-thermal fluctuations~\cite{weber2012nonthermal}.
For this reason, the field of \emph{active polymers}~\cite{winkler2020physics, winkler2017active} has gained a growing level of attention in recent years, yielding insights into the impact of non-equilibrium fluctuations and activity on structural properties of both isolated chains and suspensions of polymers~\cite{Anderson2022,zhang2023configurational,winkler2020physics, winkler2017active,kaiser2014unusual, harder2014activity, shin2015facilitation, mousavi2021active, bianco2018globulelike, foglino2019nonequilibrium, anand2018structure, anand2018structure, locatelli2021activity, philipps2022tangentially, isele2015self}. In this study, we focus on the effect of a nonequilibrium bath featuring a spatially non-uniform degree of activity on a polymeric molecule described as an ideal Rouse chain. 
While recent attention has been devoted to the impact of inhomogeneous activity~\cite{sharma2017brownian, tailleur2008statistical, schnitzer1993theory, liebchen2019optimal, caprini2022dynamics, caprini2022active, wysocki2022interacting, jahanshahi2020realization, soker2021activity, lozano2016phototaxis, vuijk2021chemotaxis, muzzeddu2022active, muzzeddu2023taxis}, there exists a notable gap in our understanding regarding its influence on polymer-like structures.
We show that nonhomogeneous active baths induce qualitatively different spatial distributions in Rouse polymers depending on their contour length and connectivity. More precisely, short polymers preferentially accumulate in regions of low bath activity, whereas long ones in regions of high bath activity. Moreover, we demonstrate that highly connected structures typically display a tendency to localize where the activity is lower. 

\smallskip
\emph{The model.}--- We study a minimal stochastic $d-$dimensional model of an ideal Rouse polymer composed by $N$ units, subjected to exponentially-correlated noises, which account for the interaction with an active bath. The chain connectivity is encoded in the matrix $M_{ij}$ \cite{sommer1995statistics}, which determine all the pairs of interacting monomers.
See Fig.~\ref{fig:sketch} for a sketch of a polymer with linear connectivity. The polymer is characterized by a quadratic Hamiltonian
\begin{equation}
    \mathcal{H}=\frac{\kappa}{2}\sum_{i=0}^{N-1}\sum_{j=0}^{N-1} M_{ij} \bm{X}_i\cdot \bm{X}_j,
\end{equation}
with $\bm{X}_i$ the position of the $i-$th monomer and $\kappa$ the coupling strength of interacting monomers. We neglect inertial effects compared to viscous forces and assume that the polymer's motion follows  the  overdamped Langevin equation
\begin{equation}
\begin{split}
    \dot {\bm{X}}_i(t)&=-\mu \nabla_{\bm{X}_i} \mathcal{H} + \mu {\rm f}_{\rm a}(\bm{X}_i)\bm{\eta}_i  + \bm \xi_i(t).
\end{split}
\label{eq:monomer_dynamics}
\end{equation}
Here, $\mu$ denotes the mobility of the monomers and $\{\bm{\xi}_i(t)\}$ are zero-mean Gaussian white noises with correlation $\left<\xi_{i\alpha}(t) \xi_{j\beta}(s)\right>=2D \delta_{ij}\delta_{\alpha \beta}\delta(t-s)$, describing thermal fluctuations. The thermal diffusivity is related to the mobility via the Einstein's relation $D=\mu T$, with the Boltzmann constant set to $k_{\rm B}=1$ throughout the paper. As a result of the collisions with the active agents dispersed in the bath, the polymer experiences additional non-thermal fluctuations which violate the detailed balance condition and drive it out of equilibrium. This effect is modeled by the active forces~$  {\rm f}_{\rm a}(\bm{X}_i)\bm{\eta}_i $ in Eq.~\eqref{eq:monomer_dynamics}, characterized by a typical magnitude which varies non-homogeneously in space according to the function ${\rm f}_{\rm a}(\bm{x})$ and are aligned with the orientation vectors~$\bm{\eta}_i$, which evolve as the Ornstein-Uhlenbeck (OU) processes $\tau \dot{\bm{\eta}}_i=-\bm{\eta}_i+\bm{\zeta}_i(t)\,$.
Here $\{\bm{\zeta}_i(t)\}$ are $N$ independent zero-mean Gaussian white noises with correlations $\left<\zeta_{i\alpha}(t) \zeta_{j\beta}(s)\right>=2\tau d^{-1}\delta_{ij}\delta_{\alpha \beta}\delta(t-s)$, and $\tau$ is the characteristic relaxation time of the OU processes which sets the persistence time of the active forces. The time-translation invariant correlation function of the orientation vectors reads $\left\langle \eta_{i\alpha}(t) \eta_{j\beta}(s) \right\rangle = \delta_{ij}\delta_{\alpha \beta} \mathcal{C}_\eta(t-s)$, with $\mathcal{C}_\eta(t-s)=d^{-1}\exp\left(-|t-s|/\tau\right)$. The variance of $\zeta_{i \alpha}$ has been chosen such that $\left< \lVert \bm{\eta}_i \rVert^2\right>=\left<  \sum_\alpha \eta_{i\alpha}^2 \right>=1$ for all $d$ and $i$.
Note that the model can alternatively be used to describe a chain of active particles, each endowed with its own polarity, which use energetic resources distributed in the bath according to the activity ${\rm f}_{\rm a}(\bm{x})$ to self-propel.

\smallskip
\emph{Effective dynamics.}--- The stochastic dynamics in Eq.~\eqref{eq:monomer_dynamics} can be rewritten within the Rouse domain~\cite{doi1988theory} for the Rouse modes $\bm{\chi}_i=\sum_{j}\varphi_{ij}\bm{X}_j$ where the matrix $\varphi_{ij}$ is chosen in such a way to diagonalize the symmetric connectivity matrix $M_{ij}$ and with the rows normalized to unity.
The modes evolve as
\begin{equation}
     \dot{\bm{\chi}}_i=-\gamma_i\bm{\chi}_i +  \sum_j \varphi_{ij}  {\rm v}(\bm{X}_j) \bm{\eta}_j +  \tilde{\bm{\xi}}_i(t),
     \label{eq:rouse_modes_dynamics}
\end{equation}
where the Gaussian white noise $\tilde{\bm{\xi}}_i(t)$ has the same statistics as  $\bm{\xi}_i(t)$, and the monomer position $\bm{X}_j$ can be rewritten in terms of the Rouse modes using the inverse transformation $\varphi^{-1}$. The typical swim speed of the monomers due to activity is ${\rm v} (\bm{x}) \equiv \mu {\rm f}_{\rm a}(\bm{x})$, which we will refer to as the activity field. The relaxation rates $\{\gamma_i\}$ of the Rouse modes in the absence of activity are proportional to the eigenvalues $\{ \lambda_i\}$ of the connectivity matrix, i.e. $\gamma_i=\gamma \lambda_i$ where $\gamma = \mu \kappa$.
Unlike the case of a Rouse polymer in a thermal bath at equilibrium, the Rouse modes are now coupled via the activity field ${\rm v}({\bf X}_i)$, which makes the analytical treatment of the problem more challenging. We denote with $\mathcal{P}(\{ \bm{\chi}\},\{ \bm{\eta}\}, t)$ the joint probability density that the Rouse modes and the orientation vectors assume the values $\{ \bm{\chi}\}$ and $\{ \bm{\eta}\}$ at time $t$, respectively.
\begin{figure*}
    \centering
    \includegraphics[width=0.98\textwidth]{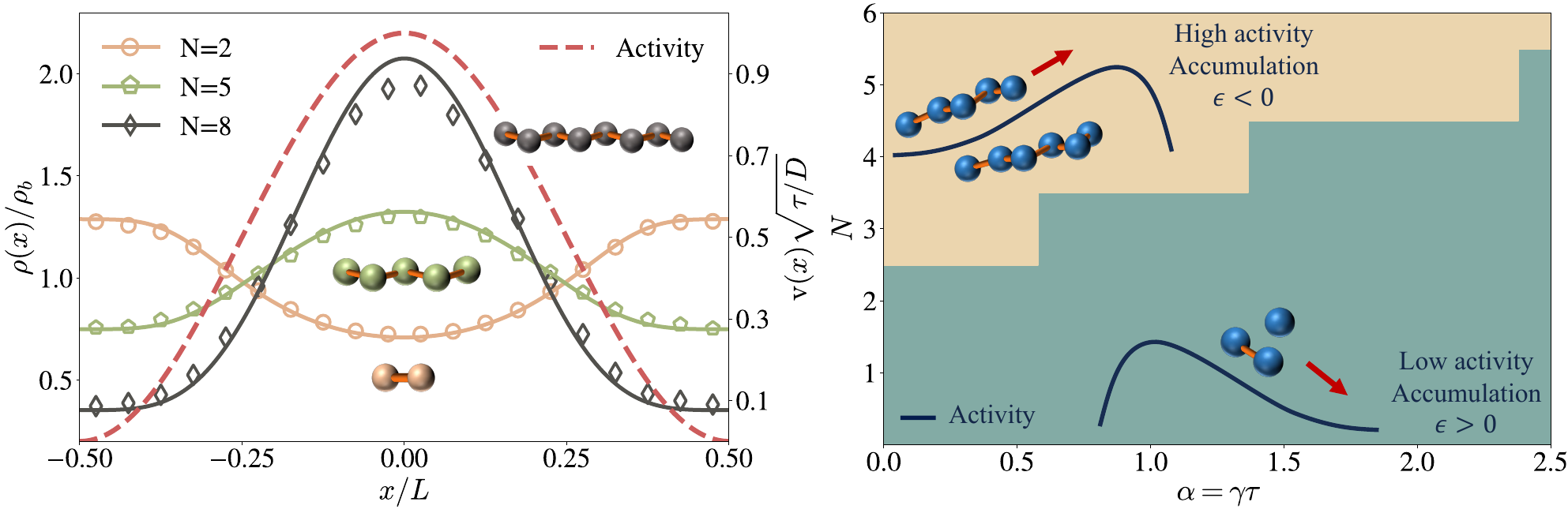}
    \caption{Left panel: 
    comparison between the analytical expression of the steady state density  $\rho$ of the polymer's center of mass (Eq.~\eqref{eq:density_com}, solid lines, left axis) and
    numerical simulations (symbols). The red dashed line shows the activity profile (right axis) and
    $\rho$ is reported in units of $\rho_b=1/L$.
    The parameters of the simulation are $T=0.1$, $\kappa=1.0$, $\mu=1.0$, $\tau=1.0$, ${\rm v}_0=1.0$, $L=10$, and the integration time step $\Delta t=0.001$. 
    Right panel: region of preferential accumulation (corresponding to the sign of $\epsilon$) of linear chains as a function of the time scale ratio $\alpha=\gamma \tau$ and the number of monomers $N$. For the purpose of visualization, the boundaries between the different signs of $\epsilon$ are drawn at half-integer values of $N$.
    }
    \label{fig:density}
\end{figure*}
Being the underlying dynamics of the system Markovian, the time evolution of $\mathcal{P}(\{ \bm{\chi}\},\{ \bm{\eta}\}, t)$ 
follows the Fokker-Planck (FP) equation~\cite{risken1996fokker, gardiner1985handbook}:
\begin{equation}
\begin{split}
    \partial_t \mathcal{P}= (\mathcal{L}_0 + \mathcal{L}_{\rm a} + \mathcal{L}_{\bm{\eta}} )   \mathcal{P},
\end{split}
\label{eq:full_FPE}
\end{equation}
with the set of operators $\{ \mathcal{L}_0, \mathcal{L}_{\rm a},  \mathcal{L}_{\bm{\eta}} \}$ defined as:
\begin{equation}
    \begin{split}
        \mathcal{L}_{0} &\equiv \sum_{i=0}^{N-1} \nabla_i \cdot \Big[ \gamma_i \bm{\chi}_i + D \nabla_i \Big],\\
        \mathcal{L}_{\rm a} &\equiv \sum_{i=0}^{N-1} \nabla_i \cdot \Big[ - \sum_j \varphi_{ij} {\rm{v}}(\bm{X}_j) \bm{\eta}_j \Big],\\
        \mathcal{L}_{\bm{\eta}} &\equiv \sum_{i=0}^{N-1} (d \tau)^{-1}\left[ \widetilde{\nabla}_i^2+ d \widetilde{\nabla}_i \cdot \bm{\eta}_i \right].
    \end{split}
    \label{eq:FP_operators}
\end{equation}
Here, we used the shorthand notation $\nabla_i \equiv \nabla_{\bm{\chi}_i}$ and $\widetilde{\nabla}_i \equiv \nabla_{\bm{\eta}_i}$. The operator $\mathcal{L}_{0}$ corresponds to the FP-operator of a free Rouse chain in contact with an equilibrium thermal bath, while the effect of the activity is brought in by $\mathcal{L}_{\rm a}$ and $\mathcal{L}_{\bm{\eta}}$.

To investigate how the spatial localization of the polymer correlates with the bath activity, we look for a description that includes the center of mass of the polymer $\bm{X}_{\rm com}=\bm{\chi}_0/\sqrt{N}$
as the only relevant variable.
Accordingly, we perform a coarse-graining procedure based on a moment expansion and a small-gradient approximation, as detailed in the Appendices~\ref{app:A} and~\ref{app:B}. In particular, we assume that the activity field ${\rm v}$ has small spatial variations on the length scales of $\ell_{\rm b}=\sqrt{dT/k}$ and $\ell_{\rm p} = {\rm v} \tau$, which correspond to the bond length and the persistence length of an active particle, respectively. 
As a consequence, the marginal density $\rho_0(\bm{\chi}_0,t)$ of the $0^\text{th}$ Rouse mode and its associated probability flux $\bm{\mathcal{J}}_{0}$ will also exhibit small gradients on the same length scales. 
This simplifying assumption makes the gradient expansion (see, e.g., Refs.~\cite{tailleur2008statistical, cates2013active, muzzeddu2023taxis}) a suitable approach to derive an effective equation for $\rho_0$.
In particular, by neglecting contributions of order $\mathcal{O}(\nabla_{0}^2)$ and higher in $\bm{\mathcal{J}}_{0}$, thus truncating the expansion to the drift/diffusion order (see Appendix~\ref{app:B} for details), we obtain that $\rho_0(\bm{\chi}_0,t)$ evolves according to
\begin{equation}
    \partial_t \rho_0=-\nabla_0 \cdot [\bm{\mathcal{V}}\rho_0 - \nabla_0 (\mathcal{D}\rho_0)]\,,
    \label{eq:effective_FP}
\end{equation}
where we introduced the effective drift $\bm{\mathcal{V}}(\bm{\chi}_0)$ and diffusivity $\mathcal{D}(\bm{\chi}_0)$ given by
\begin{eqnarray}
    \mathcal{D}(\bm{\chi}_0)&=& D + \frac{\tau}{d}\, {\rm v}^2\Big( \frac{\bm{\chi}_0}{\sqrt{N}} \Big)\,, \label{eq:effective_D}\\
    \bm{\mathcal{V}}(\bm{\chi}_0)&=&(1-\epsilon/2) \nabla_0 \mathcal{D}(\bm{\chi}_0) \,.
    \label{eq:effective_V}
\end{eqnarray}
Equation~\eqref{eq:effective_D} shows that the effective diffusivity $\mathcal{D}$ consists of the term $D$, due to thermal fluctuations, and of the enhancement induced by non-equilibrium fluctuations caused by the active forces.
Moreover, the spatial variations of the activity field induce the effective drift $\bm{\mathcal{V}}$ in Eq.~\eqref{eq:effective_V}, which is always aligned with the activity gradient.
The entity of this drift depends on the parameter $\epsilon$, which is related to the polymer architecture and to the persistence time $\tau$ of the active forces by the following expression:
\begin{equation}
    \epsilon = 1 - \sum_{i=1}^{N-1}\frac{1}{1+\tau \gamma_i}\,.
    \label{eq:epsilon}
\end{equation}
We recall here that the relaxation rates $\{ \gamma_i \}$ carry information on the polymer connectivity, being proportional to the eigenvalues $\{\lambda_i \}$ of the connectivity matrix.
In particular, for a linear chain, $\lambda_j=4 \sin^2(j \pi/2N)$.
Furthermore, it should be noted that Eq.~\eqref{eq:epsilon} gives $\epsilon<1$ for any choice of the model parameters, implying that the effective drift $\bm{\mathcal{V}}$ always points in the direction of greatest increase of the activity field.
This might lead to the wrong conclusion that all polymeric structures tend to accumulate in regions of high bath activity, driven by $\bm{\mathcal{V}}$.
However, high activity regions are also characterized by a larger effective diffusivity $\mathcal{D}$, whose effect is to reduce the typical residence time of the polymer in those areas, thus counteracting the effective drift.
The competition between these two effects is governed by $\epsilon$: depending on their degree of polymerization and connectivity, different chains will preferentially localize in different regions of space.

\smallskip
\emph{Stationary distributions.}--- In fact, by solving Eq.~\eqref{eq:effective_FP} at steady state with zero flux condition, and introducing $\rho(\bm{X}_{\rm com})\equiv \rho_0(\sqrt{N}\bm{X}_{\rm com})$, we get:
\begin{equation}
    \rho(\bm{X}_{\rm com}) = \mathcal{N} \Bigg[ 1 + \frac{\tau {\rm v}^2(\bm{X}_{\rm com})}{dD} \Bigg]^{-\epsilon/2},
    \label{eq:density_com}
\end{equation}
with $\mathcal{N}$ a normalization constant. Equation~\eqref{eq:density_com} implies that all chains with $\epsilon>0$ will typically spend more time in regions of low bath activity, whereas those with $\epsilon<0$ will preferentially accumulate in high activity areas. 

At fixed bath conditions, i.e., fixed time scale ratio $\alpha \equiv \tau \gamma = \tau \kappa \mu$, there are only two ways to vary $\epsilon$. The first one is to change the degree of polymerization $N$ of the chain by adding/removing monomeric units.
The left panel of Fig.~\ref{fig:density} shows the steady state density of the center of mass 
%reported in Eq.~\eqref{eq:density_com},
for the case of linear chains of various lengths. Theoretical predictions (Eq.~\eqref{eq:density_com}, solid lines) and numerical simulations (symbols) are compared in $d=2$, for polymers in a box with size $L$ endowed with periodic boundary conditions, and a sinusoidal activity field ${\rm v}(x)=({\rm v}_0/2)[1+\cos (2 \pi x/L)]$ along the $x$-axis and uniform along the remaining, orthogonal axis. The plot shows that short chains (e.g., dimers) preferentially localize in low-activity regions, whereas the density of longer chains increasingly peaks around regions of high activity as the number of monomers increases.
The minimum number of monomers above which linear chains localize in regions of high activity depends on the persistence time $\tau$ of the active forces and on the stiffness $\kappa$ of the interaction between interconnected monomers. 
The separation between localization in high/low activity regions is evident in the diagram of Fig.~\ref{fig:density} (right panel), which identifies the domains of the parameter space $(N,\alpha)$ corresponding to these two cases.
An immediate conclusion drawn from the diagram is that for single particles and dimers, the effective diffusivity always prevails over the drift contribution, leading to localization in regions of low activity for any value of $\alpha$. In fact, for $\alpha \to 0$,  the coefficient $\epsilon \simeq 2-N$, implying that only chains with $N>2$ localize in regions of high activity. For finite $\alpha$, the dominant contribution to the coefficient $\epsilon$ comes from Rouse modes which relax slower than the correlation time $\tau$ of the active bath. This corresponds to the observation that with increasing $\tau$, linear chains require a higher degree of polymerization to preferentially localize where the activity is larger (Fig.~\ref{fig:density}).\\
The second way to change the sign of $\epsilon$ is to vary the connectivity matrix $M_{ij}$ of the chain by keeping fixed the number $N$ of monomers.
%
%%%%%%%%%%%%%%%%%%%%%%%%%%%%%%%%%%%%%%
%%%%%%%%%%%%%%%%%%%%%%%%%%%%%%%%%%%%%%
\begin{figure}
    \centering
    \includegraphics[width=0.48\textwidth]{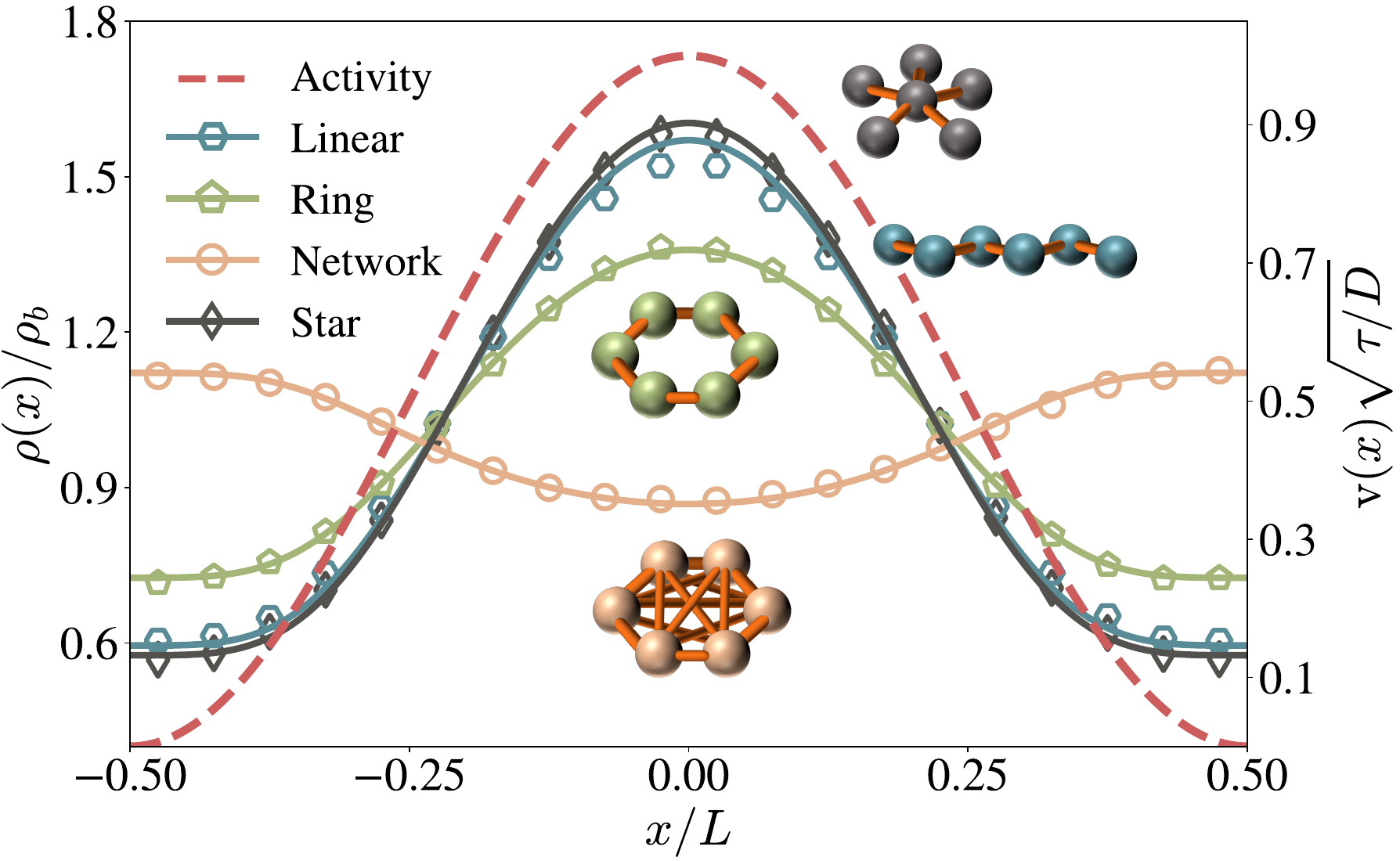}
    \caption{%The plot reports the 
    Stationary probability density $\rho$ of the center of mass of polymers with $N=6$ monomers and different connectivity matrix:  linear (blue), ring (green), star (gray), and fully connected (light orange)  polymer. We compare analytical predictions (Eq.~\eqref{eq:density_com}, solid lines) and simulations results in $d=2$ dimensions (symbols). The activity field is represented by the red dashed line (right axis).
    The parameters of the simulation are the same as in Fig.~\ref{fig:density}.
    }
    \label{fig:topology}
\end{figure}
%%%%%%%%%%%%%%%%%%%%%%%%%%%%%%%%%%%%%%
%%%%%%%%%%%%%%%%%%%%%%%%%%%%%%%%%%%%%%
%
In order to demonstrate this, we determined the stationary density $\rho(\bm{X}_{\rm com})$ for different structures, i.e. linear, ring, and star polymer as well as fully connected network. 
Figure \ref{fig:topology} shows the resulting $\rho$, obtained with the same activity field as in Fig.~\ref{fig:density}.
It turns out that, for a fixed degree $N$ of polymerization ($N=6$ in Fig.~\ref{fig:topology}), the most constrained structure from the point of view of internal interactions, i.e., the fully connected network, is unable to localize in the region of high activity, whereas the structures with a lower degree of connectivity typically spend more time where the activity is higher.
Moreover, as can be seen in Fig.~\ref{fig:topology}, the localization is more effective for those chains characterized by the least number of bonds. 
%%
%%
%%
%%%%%%%%%%%%%%%%%%%%%%%%%%%%%%%%%%%%%%%%%%%%%%%%%%%%%%%%%%%%%%%%
%%%%%%%%%%%%%%%%%%%%%%%%%%%%%%%%%%%%%%%%%%%%%%%%%%%%%%%%%%%%%%%%%
\begin{figure}[t]
    \centering
    \includegraphics[width=0.48\textwidth]{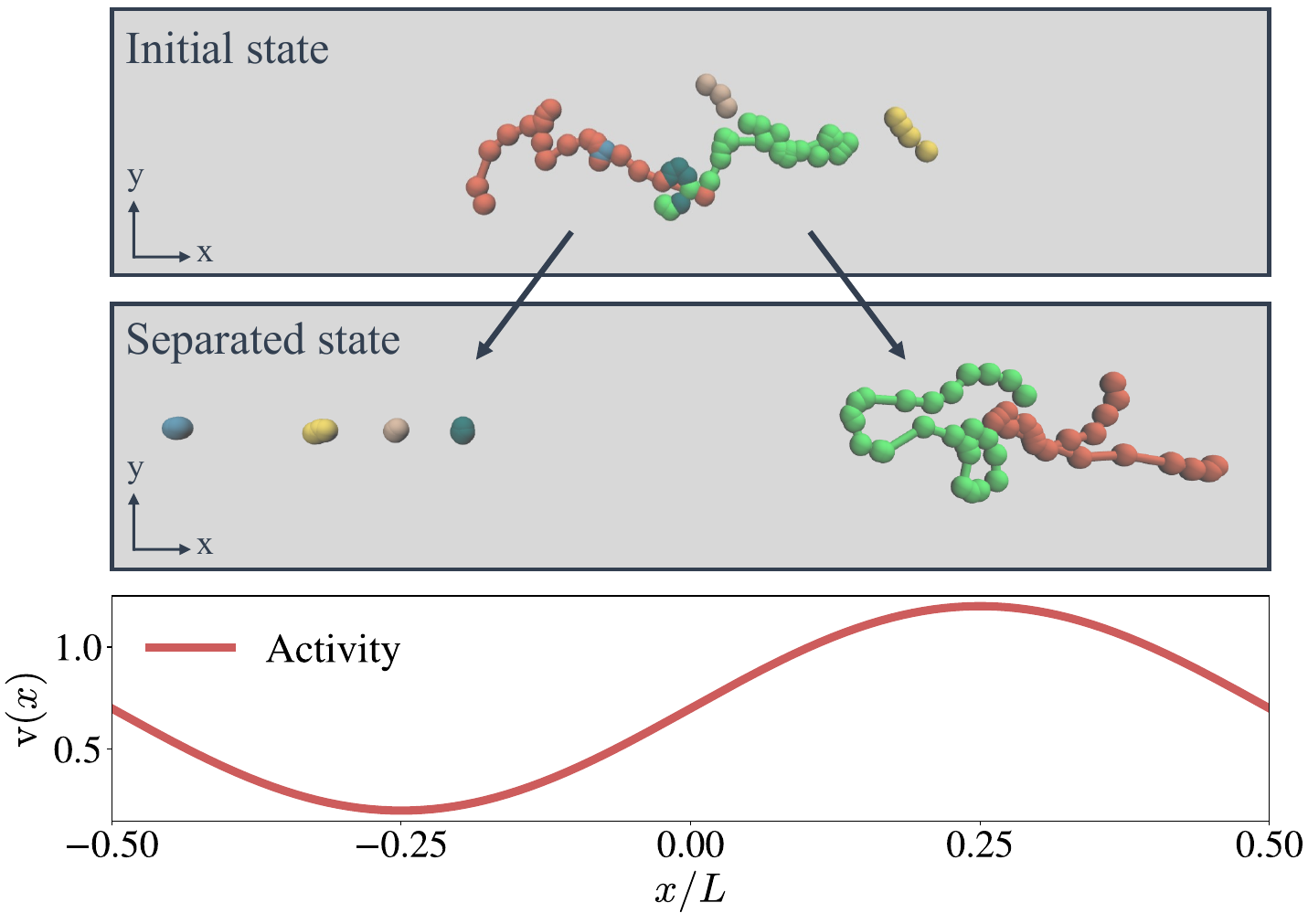}
    \caption{Molecular dynamics simulations proving the spontaneous separation of linear polymer chains with various lengths in a sinusoidal activity field ${\rm v}(x)=0.2+0.5[1+\sin (2 \pi x/L)]$ (bottom panel) and in $d=2$.
    Polymers with $N\in\{2,3,4,5,20,25\}$ are initially localized and mixed at the center of the 
    box (top snapshot). Over time, different species migrate to different regions in space, based on their polymerization degree (middle snapshot). 
    Simulation parameters: $T=0.01$, $\kappa=3.0$, $\mu=1.0$, $\tau=3.0$, $L=20$. Integration timestep: $\Delta t=0.001$. }
    \label{fig:simulations}
\end{figure}
%%%%%%%%%%%%%%%%%%%%%%%%%%%%%%%%%%%%%%%%%%%%%%%%%%%%%%%%%%%%%%%%
%%%%%%%%%%%%%%%%%%%%%%%%%%%%%%%%%%%%%%%%%%%%%%%%%%%%%%%%%%%%%%%%%
%
The fact that polymer chains localize in regions of high or low activity depending on their degree of polymerization and connectivity can lead to the spontaneous spatial separation of different polymer species, when these are immersed in a non-uniform active bath.
This can be observed, for example, in molecular dynamics simulations of a mixture of linear chains of various lengths in a sinusoidal activity field, as shown in Fig.~\ref{fig:simulations}.
After an initial phase in which all chains are prepared in a mixed phase localized around the center of the box, different species begin to migrate to different regions of space %and localize in areas of high or low activity 
according to their length.
In particular, the chains with $N=20$ and 25 localize where the activity is higher, while the shorter chains spend more time in the region of low activity.
To better appreciate the separation along the $x$-axis, where the activity is non-uniform, a harmonic confining potential along the $y$-axis (see Fig.~\ref{fig:simulations}) has been introduced.
A similar spontaneous separation, will occur even in the presence of steric hindrance and inter-chain interactions (neglected here), at least for dilute polymer mixtures, and possibly with different time scales, as the initial mixed state will take longer to untangle.

\emph{Perspectives.}--- The ability to segregate and sort biomolecules or synthetic polymer-like structures at micro/nano-meter scale is of paramount importance in a variety of applications, spanning from diagnostics and biomedicine to biological analyses and chemical processing~\cite{sajeesh2014particle}. Nonequilibrium conditions have already proved useful in length-selective accumulation of oligonucleotides subjected to thermal gradients~\cite{kreysing2015heat} and elasticity-based polymer sorting in active fluids~\cite{shin2017elasticity}. The mechanism investigated here has potential to be employed in active sorting techniques to separate polymers based on both their length and structural connectivity. 
In a more realistic setting, the relaxation time scales of the chain will be influenced by additional interactions (e.g., bending rigidity and steric hindrance) and the active bath may also exhibit more than one time scale. The ratio between the characteristic time scales of the bath and of the polymer, which determines the migration properties of the latter,
will thus be different from the case of an ideal chain discussed here.
Therefore, the transition from low activity to high activity localization will be affected, yet it will still be possible to separate polymer species based on their length or connectivity.
Moreover, our theoretical predictions might be experimentally tested with synthetic chains assembled from magnetic colloidal beads~\cite{mhanna2022chain, martinez2016orientational} immersed in a bath with photokinetic bacteria, the swimming speed of which depends on the incident light intensity~\cite{massana2022rectification, massana2024multiple}.\\

A.S.~acknowledges support by the Deutsche Forschungsgemeinschaft (DFG) within the Project No.~SH 1275/5-2. J.U.S.~thanks the cluster of excellence “Physics of Life” at TU Dresden for its support. P.L.M. thanks D. Venturelli, T. Földes, and M. Barbi for the fruitful discussions and the critical reading of the manuscript. The authors thank E. Roldan for providing feedback on the work and suggesting possible biological applications.

\bibliographystyle{apsrev4-1}
\bibliography{references}

\appendix

\section{Moment expansion}
\label{app:A}

In order to determine the spatial regions of the bath in which the chain tends to accumulate, we apply a two-step coarse-graining procedure to the dynamics of the system.
First, we marginalize the joint PDF $\mathcal{P}(\{ \bm{\chi}\},\{ \bm{\eta}\}, t)$ over the orientation vectors $\{ \bm{\eta} \}$ associated to the active forces and obtain an exact evolution equation for the marginal density
\begin{equation}
    \varrho(\{ \bm{\chi}\},t)\equiv \int \prod_{i=0}^{N-1} d\bm{\eta}_i \mathcal{P}(\{ \bm{\chi}\},\{ \bm{\eta}\}, t).
    \label{eq:marginal_definition}
\end{equation}
As a second step, under some assumptions which will be introduced in the following, we integrate out the information concerning the internal structure of the chain (i.e., the Rouse modes $\{ \bm{\chi}_i \}$ with $i>0 $)\cite{doi1988theory}, 
%
%\agc{Here put a reference to a general book where to look how to introduce the Rouse modes}
%
thus getting the marginal density $\rho_0(\bm{\chi}_0,t)$ of the rescaled center-of-mass $\bm{\chi}_0$ which is defined as
\begin{equation}
    \rho_0(\bm{\chi}_0,t)\equiv \int \prod_{i=1}^{N-1}d\bm{\chi}_i \, \varrho(\{ \bm{\chi}\},t).
    \label{eq:marginal_com}
\end{equation}
To marginalize the active degrees of freedom $\{ \bm{\eta} \}$, we first expand the joint probability density $\mathcal{P}(\{ \bm{\chi}\},\{ \bm{\eta}\}, t)$ into the eigenfunctions of the operator $\mathcal{L}_{\bm{\eta}}$ which contributes to the dynamics of $\mathcal{P}$ according to Eq.~\eqref{eq:full_FPE} of the main text. 
It can be shown that the latter is diagonalized by the following set of functions:
\begin{equation}
        u_{\bm{n}}(\left\{ \bm{\eta}\right\})=\frac{\exp \left\{ -\frac{d \sum_j\bm{\eta}_j^2}{2}\right\} \prod_{i=0}^{N-1}\prod_{\alpha=1}^d H_{n_{i \alpha}}\left( \sqrt{d} \eta_{i\alpha}\right)}{(2 \pi /d)^{Nd/2}},
        \label{eq:eigenfunctions_L}
\end{equation}
where $\bm{n}$ denotes an $N \times d$ matrix of non-negative integers used to label the eigenfunctions and $H_n(x)$ is the $n$-th Hermite polynomial in the probabilist convention~\cite{abramovitz1964handbook}. The corresponding eigenvalues $\lambda_{\bm n}$ are proportional to the inverse persistence time $1/\tau$:
\begin{equation}
    \lambda_{\bm n}=- \tau^{-1}\sum_{i=0}^{N-1} \sum_{\alpha=1}^d n_{i\alpha}\,.
    \label{eq:eigenfunvalues_L}
\end{equation}
Accordingly, the joint probability density $\mathcal{P}(\{ \bm{\chi}\},\{ \bm{\eta}\}, t)$ can be rewritten as a weighted combination of the basis elements $\{ u_{\bm{n}}(\left\{ \bm{\eta}\right\}) \}$:
\begin{equation}
    \mathcal{P}(\{ \bm{\chi}\},\{ \bm{\eta}\}, t)=\sum_{\bm n \in \mathbb{N}_0^{N \times d}} \phi_{ \bm{n}}( \left\{ \bm{\chi} \right\},t ) u_{ \bm{n}}(\left\{ \bm{\eta}\right\} )\,,
    \label{eq:expansion_joint}
\end{equation}
where the dependence on the Rouse modes and time is now brought in by the expansion coefficients $\{ \phi_{ \bm{n}}( \left\{ \bm{\chi} \right\},t ) \}$. 
In order to derive their governing equations, we find convenient to first introduce the set of auxiliary functions $\{ \tilde{u}_{\bm{n}}(\left\{ \bm{\eta}\right\})\} $ defined as:
\begin{equation}
        \tilde{u}_{\bm{n}}(\{\bm{\eta}\})= \prod_{i=0}^{N-1}\prod_{\alpha=1}^d\frac{H_{n_{i \alpha}}\left( \sqrt{d}\,\eta_{i \alpha} \right)}{n_{i \alpha}!},
        \label{eq:auxiliary_functions}
\end{equation}
which satisfy the following orthonormality relation with the eigenfunctions~\eqref{eq:eigenfunctions_L}: 
\begin{equation}
    \int \prod_{i=0}^{N-1} d \bm{\eta}_i\,\tilde{u}_{\bm{m}} u_{ \bm{n}}=\prod_{i=0}^{N-1}\prod_{\alpha=1}^d \delta_{n_{i \alpha},m_{i \alpha}}=\delta_{\bm{n},\bm{m}}.
    \label{eq:orthogonality_relation}
\end{equation}
With the help of Eqs.~\eqref{eq:expansion_joint}, \eqref{eq:auxiliary_functions}, and \eqref{eq:orthogonality_relation}, it can be easily shown that the lowest-order coefficient $\phi_{ \bm{0}}( \left\{ \bm{\chi} \right\},t )$ is nothing but the marginal distribution $\varrho(\{ \bm{\chi} \},t)$:
\begin{equation}
    \begin{split}
        \varrho(\left\{ \bm{\chi} \right\},t)&= \int \prod_{i=0}^{N-1} d \bm{\eta}_i\,\tilde{u}_{\bm{0} } \mathcal{P}( \left\{ \bm{\chi} \right\}, \left\{ \bm{\eta} \right\} ,t )\\ &= \sum_{ \bm{n}} \phi_{ \bm{n}}(\{ \bm{\chi} \},t)\int \prod_{i=0}^{N-1} d \bm{\eta}_i\,\tilde{u}_{ \bm{0} } u_{ \bm{n} }\\
        &=\phi_{\bm{0}}(\{ \bm{\chi} \},t)\,.
    \end{split}
\end{equation}
Accordingly, the information about the conformation of the polymer and its preferential accumulation in specific regions of the non-homogeneous active bath is encoded in $\phi_{\bm{0}}(\{ \bm{\chi} \},t)$. In order to derive the evolution equations for the coefficients $\{\phi_{\bm{n}}(\{ \bm{\chi} \},t)\}$, we introduce the following inner product between two generic functions $f,\,g$ of the orientation vectors $\{\bm{\eta}\}$:
\begin{equation}
    \langle f(\{\bm{\eta}\}); g(\{\bm{\eta}\}) \rangle \equiv \int \prod_{i=0}^{N-1} d\bm{\eta}_i f(\{\bm{\eta}\}) g(\{\bm{\eta}\});
\end{equation}
then, we project the FP equation~\eqref{eq:full_FPE} onto the auxiliary functions $\{ \tilde{u}_{\bm{n}}(\left\{ \bm{\eta}\right\})\} $, finding
\begin{equation}
    \partial_t \phi_{\bm{n}}(\{ \bm{\eta} \})= \langle \tilde{u}_{\bm{n}} ; \partial_t \mathcal{P} \rangle= \langle \tilde{u}_{\bm{n}} ; (\mathcal{L}_0 + \mathcal{L}_{\rm a} + \mathcal{L}_{\bm{\eta}} )   \mathcal{P} \rangle .
    \label{eq:evolution_coefficients}
\end{equation}
The right hand side of Eq.~\eqref{eq:evolution_coefficients} can be evaluated after recalling that all Hermite polynomials can be built by iteratively applying the following recurrence relation, starting from $H_0(x)=1$~\cite{abramovitz1964handbook},
\begin{equation}
        H_{n+1}(x)=xH_n(x)-H'_n(x),
        \label{eq:recurrence_Hermite}
\end{equation}
and that they form a so-called Appell sequence, as they satisfy
\begin{equation}
    H'_n(x)=nH_{n-1}(x).
    \label{eq:Appell_sequence}
\end{equation}
These two identities can be combined to obtain useful relations between Hermite polynomials of different orders. To set the notation for the upcoming derivation, we find convenient to extend the definition of $u_{\bm{n}},\, \tilde{u}_{\bm{n}} $ and $ \phi_{ \bm{n}} $ to the case with $\bm{n} \in \mathbb{Z}^{N \times d}$, assuming that $u_{\bm{n}}=\tilde{u}_{\bm{n}}=\phi_{ \bm{n}}=0$ if the matrix $\bm{n}$ contains at least one negative element. Moreover, we introduce the raising and lowering operators $b_{i \alpha}^\dagger,\,b_{i \alpha}:\,\mathbb{Z}^{N \times d} \longrightarrow \mathbb{Z}^{N \times d}$, which act on an $N \times d$ matrix $\bm{n}$ by increasing/decreasing its $(i,\alpha)$-component by a unit. With the help of Eqs.~\eqref{eq:recurrence_Hermite} and~\eqref{eq:Appell_sequence}, the following useful identity can be obtained:
\begin{equation}
        \eta_{i\alpha} H_{n_{i \alpha}} ( \sqrt{d} \eta_{i \alpha} )= \frac{ H_{n_{i \alpha}+1} ( \sqrt{d} \eta_{i \alpha} ) + n_{i\alpha} H_{n_{i \alpha}-1} ( \sqrt{d} \eta_{i \alpha} )}{\sqrt{d}},
    \label{eq:recurrence_Hermite_2}
\end{equation}
which implies:
\begin{equation}
    \eta_{i\alpha} u_{\bm{n}} =d^{-1/2} \big[u_{b^\dagger_{i\alpha}\bm{n}}+ n_{i \alpha} u_{b_{i\alpha}\bm{n}}\big].
    \label{eq:recurrence_eigenfunctions}
\end{equation}
Using the identities introduced above, we are now in the position of evaluating the right hand side of Eq.~\eqref{eq:evolution_coefficients}. To avoid cumbersome expressions, we separately determine the contributions due to the three operators $\mathcal{L}_0$, 
$\mathcal{L}_{\rm a}$, and $\mathcal{L}_{\bm{\eta}}$ given in Eq.~\eqref{eq:FP_operators}. As $\mathcal{L}_0$ does not explicitly depend on the orientation vectors $\{ \bm{\eta}_i\}$, it is straightforward to show that:
\begin{equation}
    \langle \tilde{u}_{\bm{n}} ; \mathcal{L}_0 \mathcal{P} \rangle=\sum_{\bm{m}} \langle \tilde{u}_{\bm{n}} ; u_{\bm{m}} \rangle \mathcal{L}_0 \phi_{\bm{m}}=\mathcal{L}_0 \phi_{\bm{n}}.
    \label{eq:proj_L0}
\end{equation}
The projection of $\mathcal{L}_{\bm{\eta}} \mathcal{P}$ onto $\tilde{u}_{\bm{n}}$ can be easily computed by exploiting the fact that $\mathcal{L}_{\bm{\eta}}$ does not depend on the Rouse modes and it is diagonalized by the eigenfunctions $\{u_{\bm{n}}\}$:
\begin{equation}
    \langle \tilde{u}_{\bm{n}} ; \mathcal{L}_{\bm{\eta}} \mathcal{P} \rangle=\sum_{\bm{m}} \phi_{\bm{m}} \langle \tilde{u}_{\bm{n}} ; \mathcal{L}_{\bm{\eta}} u_{\bm{m}} \rangle=\lambda_{\bm{n}}\phi_{\bm{n}}.
    \label{eq:proj_Leta}
\end{equation}
Deriving the contribution coming from the projection of $\mathcal{L}_{\rm a} \mathcal{P}$ is slightly more complicate, as it contains information about the coupling between the Rouse modes and the orientation vectors. It reads
\begin{equation}
    \begin{split}
        \langle \tilde{u}_{\bm{n}} ; \mathcal{L}_{\rm a} \mathcal{P} \rangle&= -\partial_{i \alpha} \varphi_{ij} {\rm{v}}(\bm{X}_j) \sum_{\bm{m}} \phi_{\bm{m}} \langle \tilde{u}_{\bm{n}} ; \eta_{j \alpha} u_{\bm{m}} \rangle \\
        &=-\partial_{i \alpha} \varphi_{ij} {\rm{v}}(\bm{X}_j) \frac{1}{\sqrt{d}}[\phi_{b_{j \alpha}\bm{n}}+(n_{j \alpha}+1)\phi_{b^\dagger_{j \alpha}\bm{n}}],
    \end{split}
    \label{eq:proj_La}
\end{equation}
where summation over repeated indices is implied and we used Eq.~\eqref{eq:recurrence_eigenfunctions} in order to evaluate the inner product in the right hand side of the first line:
\begin{equation}
    \begin{split}
        \sqrt{d}\, \langle \tilde{u}_{\bm{n}} ; \eta_{j \alpha} u_{\bm{m}} \rangle&=\langle \tilde{u}_{\bm{m}} ; u_{b^\dagger_{j\alpha}\bm{m}} \rangle + n_{j \alpha} \langle \tilde{u}_{\bm{m}} ;  u_{b_{j\alpha}\bm{m}} \rangle\\
        &=\delta_{\bm{n},b^\dagger_{j\alpha}\bm{m}} + m_{j \alpha} \delta_{\bm{n},b_{j\alpha}\bm{m}}\\
        &=\delta_{b_{j\alpha}\bm{n},\bm{m}} + (n_{j \alpha}+1) \delta_{b^\dagger_{j\alpha}\bm{n},\bm{m}}.
    \end{split}
    \label{eq:identity_inn_prod}
\end{equation}
The projection in Eq.~\eqref{eq:evolution_coefficients} generates a system of coupled partial differential equations for $\{\phi_{\bm{n}}(\{ \bm{\chi} \},t)\}$ with a hierarchical structure 
\begin{equation}
    \begin{split}
        \partial_t \phi_{\bm{n}}&=\mathcal{L}_0 \phi_{\bm{n}} + \lambda_{\bm{n}}\phi_{\bm{n}} \\
        &-\partial_{i \alpha} \varphi_{ij} {\rm{v}}(\bm{X}_j) \frac{1}{\sqrt{d}}[\phi_{b_{j \alpha}\bm{n}}+(n_{j \alpha}+1)\phi_{b^\dagger_{j \alpha}\bm{n}}]\,,
    \end{split}
    \label{eq:evolution_coefficients_full}
\end{equation}
which bears similarities to hydrodynamic theories.
For this reason, Eq.~\eqref{eq:evolution_coefficients_full} is sometimes referred to as a generalized hydrodynamic hierarchy~\cite{marchetti2013hydrodynamics, cates2013active}, even though our model neglects any explicit hydrodynamic effect due to the interaction of the polymer with the surrounding fluid. 

It is possible to show that the expansion coefficients $\{ \phi_{ \bm{n}}( \left\{ \bm{\chi} \right\},t ) \}$ are related to the conditional moments of the orientation vectors given a fixed polymer configuration $\left\{ \bm{\chi} \right\}$. In fact, the following equalities hold:
\begin{equation}
    \begin{split}
        \varrho&=\varrho \langle 1 | \left\{ \bm{\chi} \right\} \rangle  = \phi_{\bm{0}} \,,\\ 
        \sigma_{i \alpha} &\equiv \varrho \langle \eta_{i \alpha} | \left\{ \bm{\chi} \right\} \rangle =  \frac{1}{\sqrt{d}}\phi_{b^\dagger_{i \alpha}\bm{0}}\,,\\ 
        Q_{i j \alpha \beta} &\equiv \varrho \langle \eta_{i\alpha} \eta_{j\beta} - \frac{\delta_{ij}\delta_{\alpha \beta}}{d}  |  \left\{ \bm{\chi} \right\} \rangle \\&= d^{-1}[1 + \delta_{ij}\delta_{\alpha \beta}]\phi_{b^\dagger_{i \alpha} b^\dagger_{j \beta} \bm{0}}\,,
    \end{split}
    \label{eq:conditional_moments}
\end{equation}
%
%\agc{I think that the $\varrho$ on the rhs of the first equation is a misprint, right?}
%
with $\bm{0}$ the $N \times d$ matrix with all entries equal to zero, and where we introduced the rank-$2$ tensor $\sigma_{i \alpha}$ and rank-$4$ tensor $Q_{i j \alpha \beta}$.
Analogous formulas which relate higher-order conditional moments and expansion coefficients can be derived.
It is worth noting that in the case $N=1$ of a single particle, the formal definitions of the newly introduced quantities $\bm{\sigma}(\{ \bm{\chi}\},t)$ and $\bm{Q}(\{ \bm{\chi}\},t)$ is analogous to those of local polar and nematic order parameters commonly employed in other contexts~\cite{de1993physics, chaikin1995principles, marchetti2013hydrodynamics, kalz2023fieldtheory}.

The evolution of the zeroth- and first-order expansion coefficients can be obtained by specializing Eq.~\eqref{eq:evolution_coefficients_full} to $\bm{n}=\bm{0}$ and $\bm{n}=b^\dagger_{i \alpha}\bm{0}$, respectively, and is given by:
\begin{equation}
    \begin{split}
        \partial_t \varrho&=-\partial_{i \alpha} \Big[ -\gamma_i \chi_{i \alpha} \varrho - D \partial_{i \alpha} \varrho +  \varphi_{ij} {\rm{v}}(\bm{X}_j) \sigma_{j \alpha} \Big]  \,, \\
         \partial_t \sigma_{i\alpha}&=  -\partial_{l \beta} \Big[ \varphi_{lj} {\rm{v}}(\bm{X}_j)  \Big(\frac{\delta_{ij}\delta_{\alpha \beta}}{d} \varrho + Q_{ji \alpha \beta} \Big) \Big] \\ &\qquad +\mathcal{L}_0 \sigma_{i \alpha}- \tau^{-1}\sigma_{i\alpha}\,.
    \end{split}
    \label{eq:rho_sigma_evolution}
\end{equation}
where we recall that $\bm{X}_j=\sum_k\varphi_{jk}^{-1} \bm{\chi}_k$. 
A few remarks on these equations are needed.
The marginal density $\varrho$ is a locally conserved quantity, hence its evolution takes the form of a continuity equation $\partial_t \varrho = -\partial_{i \alpha} J_{i \alpha}$, where
\begin{equation}
    J_{i \alpha}=-\gamma_i \chi_{i \alpha} \varrho +  \varphi_{ij} {\rm{v}}(\bm{X}_j) \sigma_{j \alpha}  - D \partial_{i \alpha} \varrho 
    \label{eq:fluxes_rouse_space}
\end{equation}
denotes the $(i,\alpha)$-component of the probability flux in the  $N\times d$ dimensional Rouse space.
The first two terms on the right hand side of Eq.~\eqref{eq:fluxes_rouse_space} correspond to the drift component of the flux $J_{i \alpha}$.
They originate from the internal interactions along the polymer backbones and the average polarity of the active forces, respectively.
On the other hand, the last term on the right hand side of Eq.~\eqref{eq:fluxes_rouse_space} arises from fluctuations due to thermal diffusion.
It can be readily shown that the marginal densities $\rho_i(\bm{\chi}_i,t)$ defined as
\begin{equation}
    \rho_i(\bm{\chi}_i,t)\equiv \int \prod_{j\neq i}d\bm{\chi}_j \,\varrho(\{ \bm{\chi}\},t),
    \label{eq:marginal_modes}
\end{equation}
also evolve according to continuity equations $\partial_t \rho_i = -\partial_{\alpha} \mathcal{J}_{i\alpha}$, with flux
\begin{equation}
\begin{split}
    \mathcal{J}_{i\alpha} &\equiv \int \prod_{h\neq i}d\bm{\chi}_h J_{i \alpha}\\&=-\gamma_i \chi_{i \alpha} \rho_i +  \varphi_{ij} \int \prod_{h\neq i}d\bm{\chi}_h {\rm{v}}(\bm{X}_j) \sigma_{j \alpha} - D \partial_{i \alpha} \rho_i \,.
    \end{split}
    \label{eq:flux_marginal_mode}
\end{equation}
Due to the fact that the probability density is locally conserved, both $\varrho$ and $\{ \rho_i \}$ can be seen as slow modes of the generalized hydrodynamic theory, i.e., they exhibit a slow relaxation when subject to large-wavelength perturbations.
On the contrary, $\sigma_{i \alpha}$ does not obey any conservation law and its relaxation occurs on a typical time scale given by the persistence time $\tau$, even when perturbed on very large length scales. For this reason, it is identified as a fast mode.
Analogously, all equations governing the evolution of the higher-order modes associated to the expansion coefficients $\phi_{\bm{n}}$ will be characterized by a damping
%
%\agc{Check if you like this change; perhaps we should add a minus sign below}
%
term $\propto \lambda_{\bm{n}}\phi_{\bm{n}}$ [see Eq.~\eqref{eq:evolution_coefficients_full}], and thus by an inverse relaxation time $|\lambda_{\bm{n}}|$.

\section{Small gradient approximation}
\label{app:B}
In this Appendix we derive the effective FP equation for the marginal probability density $\rho_0(\bm{\chi}_0,t)$ by applying a gradient expansion approach up to the drift/diffusion order (i.e., we neglect terms of order $\mathcal{O}(\nabla_0^2)$ or higher in the flux $\mathcal{J}_{0\alpha}$). We start by considering the evolution of the tensor $Q_{ij\alpha \beta}$ obtained combining Eqs.~\eqref{eq:conditional_moments} and~\eqref{eq:evolution_coefficients_full}:
\begin{equation}
\begin{split}
    \partial_t Q_{ij\alpha \beta} &= \mathcal{L}_0 Q_{ij\alpha \beta} - (2/\tau) Q_{ij\alpha \beta} \\ &\quad -\partial_{k\gamma} \varphi_{kh} {\rm v}(\bm{X}_h)d^{-1}[\delta_{hi}\delta_{\gamma \alpha} \sigma_{j\beta}+\delta_{hj}\delta_{\gamma \beta} \sigma_{i\alpha}]\\
    &\quad -\partial_{k\gamma} \varphi_{kh} {\rm v}(\bm{X}_h) \Sigma_{ijh\alpha \beta \gamma}\,,
\end{split}
\label{eq:evolution_Q}
\end{equation}
where we used $\lambda_{ b^\dagger_{i \alpha} b^\dagger_{j \beta} \bm{0}}=-2/\tau$, the identity
\begin{equation}
    (1 + \delta_{ij}\delta_{\alpha \beta}) \phi_{b_{h\gamma}b^\dagger_{i \alpha} b^\dagger_{j \beta} \bm{0}}=d^{1/2}[\delta_{hi}\delta_{\gamma \alpha} \sigma_{j\beta}+\delta_{hj}\delta_{\gamma \beta} \sigma_{i\alpha}]\,,
    \label{eq:identity_C1}
\end{equation}
and we introduced the tensor $\Sigma_{ijh\alpha \beta \gamma}$ related to the third order expansion coefficient $\phi_{b^\dagger_{h\gamma}b^\dagger_{i \alpha} b^\dagger_{j \beta} \bm{0}}$, the expression of which is actually not needed below. Imposing $\partial_tQ_{ij\alpha \beta}=0$ due to the time-scale separation between slow and fast modes, we get the following compact equation for $Q_{ij\alpha \beta}$:
\begin{equation}
    Q_{ij\alpha \beta}=\partial_{k\gamma} \Upsilon_{ijk\alpha \beta \gamma},
    \label{eq:quasi_static_Q}
\end{equation}
where $\Upsilon_{ijk\alpha \beta \gamma}=\Upsilon_{ijk\alpha \beta \gamma}[\bm{\sigma}, \bm{Q},...]$ is a functional of all fast modes. This means that the tensor $Q_{ij\alpha \beta}$ is a quantity of order $\mathcal{O}(\{ \nabla_k\})$ or higher, where the notation $\mathcal{O}(\{ \nabla_k\})$ indicates dependence on first-order gradients with respect to all Rouse modes $\{ \bm{\chi}_k \}$. 
Note, however, that our gradient expansion is based on the assumption that the probability densities and fluxes (and thus the modes) exhibit small variations when the center-of-mass of the polymer 
is displaced, but not when its internal structure is changed. 
Combining Eqs.~\eqref{eq:quasi_static_Q} and~\eqref{eq:rho_sigma_evolution}, and applying again the quasistatic approximation $\partial_t \sigma_{i\alpha}=0$, we obtain the following equation for $\sigma_{i \alpha}$:
\begin{equation}
\begin{split}
    \sigma_{i \alpha}=&-\tau/d\, \partial_{l\alpha} \varphi_{li}{\rm v}(\bm{X}_i) \varrho \\
    &\quad +\tau \partial_{j\beta}\gamma_j \chi_{j\beta}\sigma_{i \alpha}\\
    &\quad - \tau \partial_{l\beta} \varphi_{lj}{\rm v}(\bm{X}_j) \partial_{k\gamma} \Upsilon_{ijk\alpha \beta \gamma}\\
    &\quad +\tau D \partial_{j\beta} \partial_{j\beta}\sigma_{i \alpha}.
\end{split}
\label{eq:quasi_stationary_sigma}
\end{equation}
Let us now consider the probability flux $\mathcal{J}_{0\alpha}$ related to the marginal density $\rho_0(\bm{\chi}_0,t)$, the definition of which is given in Eq.~\eqref{eq:flux_marginal_mode}. In particular, we focus on the second contribution on the right hand side of that equation, which originates from the interaction of the polymer with the active bath:
\begin{equation}
    \mathcal{J}^{\rm act}_{0 \alpha} \equiv \varphi_{0i} \int \prod_{h\neq 0}d\bm{\chi}_h {\rm{v}}(\bm{X}_i) \sigma_{i \alpha},
    \label{eq:flux_activity_contribution}
\end{equation}
and combine it with the expression of the quasistatic mode $\sigma_{i\alpha}$ in Eq.~\eqref{eq:quasi_stationary_sigma}. The last line of Eq.~\eqref{eq:quasi_stationary_sigma} contributes to the flux with a term proportional to:
\begin{equation}
    \begin{split}
        &\varphi_{0i} \int \prod_{h\neq 0}d\bm{\chi}_h {\rm{v}}(\bm{X}_i)  \partial_{j\beta} \partial_{j\beta}\sigma_{i \alpha}\\&=  \varphi_{0i} \int \prod_{h\neq 0}d\bm{\chi}_h [{\rm{v}}(\bm{X}_i)  \nabla_0^2 \sigma_{i \alpha} + \sigma_{i \alpha} \sum_{j\neq 0} \nabla_j^2 {\rm{v}}(\bm{X}_i)]\\&=\mathcal{O}(\nabla_0^2),
    \end{split}
    \label{eq:2nd_order_term_C1}
\end{equation}
where we used integration by parts from first to second line, and the following identity
\begin{equation}
\begin{split}
    \partial_{j\beta} {\rm{v}}(\bm{X}_i)&= \partial_{j\beta} {\rm{v}}(\varphi_{ki}\bm{\chi}_k)=\frac{\partial {\rm{v}}}{\partial \varphi_{ki}\chi_{k\alpha}} \partial_{j \beta} \varphi_{li}\chi_{l\alpha}\\&=\sqrt{N}\varphi_{ji}\partial_{0\beta}{\rm{v}} (\bm{X}_i),
    \end{split}
    \label{eq:identity_C2}
\end{equation}
to show that each Laplacian $\nabla_j^2{\rm{v}}(\bm{X}_i)$ in Eq.~\eqref{eq:2nd_order_term_C1} can be also rewritten as a term of order $\mathcal{O}(\nabla_0^2)$. Hence, the contribution to the flux $\mathcal{J}^{\rm act}_{0 \alpha}$ related to Eq.~\eqref{eq:2nd_order_term_C1} is negligible under the assumptions we made. Analogously, also the third line of Eq.~\eqref{eq:quasi_stationary_sigma} can be shown to produce terms of order $\mathcal{O}(\nabla_0^2)$ once plugged into Eq.~\eqref{eq:flux_activity_contribution}, due to
\begin{equation}
    \begin{split}
        &\int \prod_{h\neq 0}d\bm{\chi}_h {\rm{v}}(\bm{X}_i)  \partial_{l\beta} {\rm v}(\bm{X}_j) \partial_{k\gamma} \Upsilon_{ijk\alpha \beta \gamma}\\&=\mathcal{O}(\nabla_0^2)-\sum_{k\neq 0}\int \prod_{h\neq 0}d\bm{\chi}_h \Upsilon_{ijk\alpha \beta \gamma} \partial_{k \gamma } {\rm{v}}(\bm{X}_i)  \partial_{l\beta} {\rm v}(\bm{X}_j)\\&=\mathcal{O}(\nabla_0^2),
    \end{split}
    \label{eq:2nd_order_term_C2}
\end{equation}
where we used again integration by parts and Eq.~\eqref{eq:identity_C2}. This implies that the information about higher-order modes is not relevant if we are only interested in deriving an effective drift/diffusion equation for $\rho_0$. 
%
%
% QUIQUIQUIQUQIUI
%
We now focus on the terms of Eq.~\eqref{eq:quasi_stationary_sigma} that lead to non-vanishing contributions to the flux $\mathcal{J}^{\rm act}_{0 \alpha}$. Plugging the first line of Eq.~\eqref{eq:quasi_stationary_sigma} into Eq.~\eqref{eq:flux_activity_contribution} we get:
\begin{equation}
    \mathcal{J}^{{\rm act},1}_{0\alpha} \equiv -\tau/d\,\varphi_{0i} \int \prod_{h\neq 0}d\bm{\chi}_h {\rm{v}}(\bm{X}_i)\partial_{l\alpha} \varphi_{li}{\rm v}(\bm{X}_i) \varrho\,.
    \label{eq:Jact_1}
\end{equation}
We find convenient to divide the implicit sum over the index $l$ in this expression into the terms with $l=0$ and $l\neq 0$. For $l\neq0$, we get:
\begin{equation}
\begin{split}
    &-(\tau/d)\varphi_{0i} \sum_{l \neq 0}\int \prod_{h\neq 0}d\bm{\chi}_h {\rm{v}}(\bm{X}_i)\partial_{l\alpha} \varphi_{li}{\rm v}(\bm{X}_i) \varrho\\&=
    \frac{\tau}{2d}\,\varphi_{0i} \sum_{l \neq 0}\int \prod_{h\neq 0}d\bm{\chi}_h \sqrt{N} \varphi_{li}\varphi_{li} \varrho \, \partial_{0\alpha}{\rm{v}}^2(\bm{X}_i)\\&=
    \frac{\tau}{2d}\, \partial_{0\alpha}{\rm{v}}^2\Big(\frac{\bm{\chi}_0}{\sqrt{N}}\Big) \sum_{l \neq 0}\sqrt{N}\varphi_{0i} \varphi_{li}\varphi_{li} \rho_0 + \mathcal{O}(\nabla_0^2)\\&=\frac{(N-1)\tau}{2d}\rho_0 \partial_{0\alpha}{\rm{v}}^2\Big(\frac{\bm{\chi}_0}{\sqrt{N}}\Big) + \mathcal{O}(\nabla_0^2),
\end{split}
\label{eq:Jact_lnot0}
\end{equation}
where the first equality follows from integration by parts and Eq.~\eqref{eq:identity_C2}, the second equality is obtained by 
Taylor expanding $\partial_{0\alpha}{\rm{v}}^2(\bm{X}_i)$ as
\begin{equation}
    \partial_{0\alpha}{\rm{v}}^2(\bm{X}_i)=\partial_{0\alpha}{\rm{v}}^2(\varphi_{ji}\bm{\chi}_j)=\partial_{0\alpha} {\rm{v}}^2(\varphi_{0i}\bm{\chi}_0)+\mathcal{O}(\nabla_0^2),
    \label{eq:taylor_activity}
\end{equation}
and by using the definition of $\rho_0$ given in Eq.~\eqref{eq:marginal_com}. The last line of Eq.~\eqref{eq:Jact_lnot0}, instead, results from using the following identity
\begin{equation}
    \sum_{i} \sum_{l\neq 0} \varphi_{0i}\varphi_{li}\varphi_{li} =\sum_{l\neq0}\delta_{ll}/\sqrt{N}=(N-1)/\sqrt{N},
    \label{eq:identity_C3}
\end{equation}
based on the fact that the matrix $\varphi_{ij}$ is orthogonal, i.e., $\varphi_{ji}=\varphi^{-1}_{ij}$, and the entries of its first row are $\varphi_{0i}=N^{-1/2}$ for all values of $i \in \{0,1,...,N-1\}$.
For $l=0$, instead, Eq.~\eqref{eq:Jact_1} reads
\begin{equation}
\begin{split}
    &-(\tau/d)\varphi_{0i} \int \prod_{h\neq 0}d\bm{\chi}_h {\rm{v}}(\bm{X}_i)\partial_{0\alpha} \varphi_{0i}{\rm v}(\bm{X}_i) \varrho\\&=-\frac{\tau}{2d}\rho_0 \partial_{0\alpha}{\rm v}^2\Big(\frac{\bm{\chi}_0}{\sqrt{N}}\Big) - \frac{\tau}{d}{\rm v}^2\Big(\frac{\bm{\chi}_0}{\sqrt{N}}\Big)\partial_{0\alpha}\rho_0 + \mathcal{O}(\nabla_0^2)\,.
\end{split}
\label{eq:Jact_l0}
\end{equation}
Combining Eq.~\eqref{eq:Jact_lnot0} and Eq.~\eqref{eq:Jact_l0} one has:
\begin{equation}
\begin{split}
    \mathcal{J}^{{\rm act},1}_{0\alpha}&=\frac{(N-2)\tau}{2d}\rho_0 \partial_{0\alpha}{\rm{v}}^2\Big(\frac{\bm{\chi}_0}{\sqrt{N}}\Big) \\
    & \quad- \frac{\tau}{d}{\rm v}^2\Big(\frac{\bm{\chi}_0}{\sqrt{N}}\Big)\partial_{0\alpha}\rho_0 + \mathcal{O}(\nabla_0^2).
    \end{split}
    \label{eq:Jact_total}
\end{equation}
Finally, the last contribution to the flux $\mathcal{J}^{\rm act}_{0 \alpha}$ comes from inserting the second line of Eq.~\eqref{eq:quasi_stationary_sigma} into Eq.~\eqref{eq:flux_activity_contribution}.
We denote this contribution by $\mathcal{J}^{{\rm act},2}_{0\alpha}$, the expression of which is
\begin{equation}
\begin{split}
    \mathcal{J}^{{\rm act},2}_{0\alpha}&=\sum_{j}\mathcal{I}_{j\alpha}\,,\\
    \mathcal{I}_{j\alpha}&\equiv -\tau \varphi_{0i} \int \prod_{h\neq 0}d\bm{\chi}_h [\partial_{j\beta}{\rm{v}}(\bm{X}_i)] \gamma_j \chi_{j\beta}\sigma_{i \alpha},
\end{split}
\label{eq:Jact_2}
\end{equation}
where the right hand side of the second line is not implicitly summed over $j$. Using again Eq.~\eqref{eq:quasi_stationary_sigma} and neglecting all terms of order $\mathcal{O}(\nabla_0^2)$, we can derive a self-consistent equation for $\mathcal{I}_{j\alpha}$. As a first step we rewrite $\mathcal{I}_{j\alpha}$ as:
\begin{equation}
\begin{split}
    \mathcal{I}_{j\alpha}&=\frac{\tau^2}{d} \, \varphi_{0i} \int \prod_{h\neq 0}d\bm{\chi}_h [\partial_{j\beta}{\rm{v}}(\bm{X}_i)] \gamma_j \chi_{j\beta} \partial_{l\alpha} \varphi_{li}{\rm v}(\bm{X}_i) \varrho\\& -\tau^2 \varphi_{0i} \int \prod_{h\neq 0}d\bm{\chi}_h [\partial_{j\beta}{\rm{v}}(\bm{X}_i)] \gamma_j \chi_{j\beta}  \partial_{l\gamma}\gamma_l \chi_{l\gamma}\sigma_{i \alpha}\\&+\mathcal{O}(\nabla_0^2).
\end{split}
\label{eq:self_consistent_I}
\end{equation}
The first line can be simplified as follows:
\begin{equation}
\begin{split}
    &\frac{\tau^2}{d} \, \varphi_{0i} \int \prod_{h\neq 0}d\bm{\chi}_h [\partial_{j\beta}{\rm{v}}(\bm{X}_i)] \gamma_j \chi_{j\beta} \partial_{l\alpha} \varphi_{li}{\rm v}(\bm{X}_i) \varrho\\&=-\frac{\tau^2}{d} \, \varphi_{0i} \int \prod_{h\neq 0}d\bm{\chi}_h [\partial_{j\alpha}{\rm{v}}(\bm{X}_i)] \gamma_j \varphi_{ji}{\rm v}(\bm{X}_i) \varrho +\mathcal{O}(\nabla_0^2)\\& =-\frac{\tau^2}{d} \, \sqrt{N}\varphi_{0i}\varphi_{ji}\varphi_{ji} \gamma_j \rho_0 {\rm{v}}^2\Big(\frac{\bm{\chi}_0}{\sqrt{N}}\Big)+\mathcal{O}(\nabla_0^2)\\&=-\frac{\tau^2 \gamma_j}{d} \,  \rho_0 {\rm{v}}^2 \Big(\frac{\bm{\chi}_0}{\sqrt{N}}\Big)+\mathcal{O}(\nabla_0^2),
\end{split}
\label{eq:I_C1}
\end{equation}
where we used integration by parts and Eqs.~\eqref{eq:identity_C2}, \eqref{eq:taylor_activity}, and \eqref{eq:identity_C3}. The second line of Eq.~\eqref{eq:self_consistent_I} can be rewritten as:
\begin{equation}
\begin{split}
    &-\tau^2 \varphi_{0i} \int \prod_{h\neq 0}d\bm{\chi}_h [\partial_{j\beta}{\rm{v}}(\bm{X}_i)] \gamma_j \chi_{j\beta}  \partial_{l\gamma}\gamma_l \chi_{l\gamma}\sigma_{i \alpha}\\&=\tau^2 \varphi_{0i} \int \prod_{h\neq 0}d\bm{\chi}_h [\partial_{j\beta}{\rm{v}}(\bm{X}_i)] \gamma_j \gamma_j \chi_{j\beta}\sigma_{i \alpha}+\mathcal{O}(\nabla_0^2)\\&=-\tau \gamma_j \mathcal{I}_{j \alpha} +\mathcal{O}(\nabla_0^2)\,.
\end{split}
\label{eq:I_C2}
\end{equation}
Equations \eqref{eq:I_C1}, \eqref{eq:I_C2}, \eqref{eq:self_consistent_I}, and \eqref{eq:Jact_2} can eventually be combined to get\\
\begin{equation}
\begin{split}
    \mathcal{J}^{{\rm act},2}_{0\alpha}&=\sum_{j}\mathcal{I}_{j\alpha}\\&=-\frac{\tau}{2d} \Bigg[\sum_{j=0}^{N-1}\frac{\tau \gamma_j}{1+\tau \gamma_j}\Bigg] \rho \partial_{0\alpha}{\rm{v}}^2\Big(\frac{\bm{\chi}_0}{\sqrt{N}}\Big) + \mathcal{O}(\nabla_0^2)\\&=-\frac{\tau}{2d} [N-2+\epsilon ] \rho \partial_{0\alpha}{\rm{v}}^2\Big(\frac{\bm{\chi}_0}{\sqrt{N}}\Big) + \mathcal{O}(\nabla_0^2),
\end{split}
\label{eq:Jact_2_expression}
\end{equation}
where $\epsilon$ is defined in Eq.~\eqref{eq:epsilon} of the main text. Putting together the contributions derived in Eqs.~\eqref{eq:Jact_1} and~\eqref{eq:Jact_2_expression}, we can finally write down the expression of the probability flux $\mathcal{J}_{0 \alpha}$ related to the marginal density $\rho_0$, given by
\begin{equation}
\begin{split}
    \mathcal{J}_{0 \alpha}&=\mathcal{J}^{{\rm act},1}_{0 \alpha}+\mathcal{J}^{{\rm act},2}_{0 \alpha}-D\partial_{0\alpha}\rho_0\\[2mm]
    &=-\frac{\tau \epsilon}{2d}   \rho_0 \partial_{0\alpha}{\rm{v}}^2\Big(\frac{\bm{\chi}_0}{\sqrt{N}}\Big) -\Big[D+\frac{\tau}{d}{\rm v}^2\Big(\frac{\bm{\chi}_0}{\sqrt{N}}\Big)\Big]\partial_{0\alpha}\rho_0 \\
    & \qquad + \mathcal{O}(\nabla_0^2).
\end{split}
\label{eq:Jact_final_expression}
\end{equation}
After applying the chain rule to the second line of Eq.~\eqref{eq:Jact_final_expression}, we can identify the drift and diffusion component of the flux with:
\begin{equation}
    \begin{split}
        \bm{\mathcal{J}}^{{\rm drift}}_{0}&= \frac{\tau (2-\epsilon)}{2d}   \rho_0 \nabla_0{\rm{v}}^2\Big(\frac{\bm{\chi}_0}{\sqrt{N}}\Big),\\[2mm]
        \bm{\mathcal{J}}^{{\rm diff}}_{0}&=-\nabla_0\Big(\Big[D+\frac{\tau}{d}{\rm v}^2\Big(\frac{\bm{\chi}_0}{\sqrt{N}}\Big)\Big]\rho_0 \Big),
    \end{split}
    \label{eq:drift_diffusion_fluxes}
\end{equation}
from which Eq.~\eqref{eq:effective_FP} of the main text follows.

\end{document}